# SOA - NOLM in Reflective Configuration for Optical Regeneration in High Bit Rate Transmission Systems

V. Roncin, G. Girault, M. Gay, L. Bramerie and J. C. Simon

*Abstract*— T This paper presents a theoretical and experimental investigation of optical signal regeneration properties of a non-linear optical loop mirror using a semiconductor optical amplifier as the active element (SOA-NOLM). While this device has been extensively studied for optical time division demultiplexing (OTDM) and wavelength conversion applications, our proposed approach, based on a reflective configuration, has not yet been investigated, particularly in the light of signal regeneration. The impact on the transfer function shape of different parameters, like SOA position in the interferometer and SOA input optical powers, are numerically studied to appreciate the regenerative capabilities of the device. Regenerative performances in association with a dual stage of SOA to create a 3R regenerator which preserves the data polarity and the wavelength are experimentally assessed. Thanks to this complete regenerative function, a 100.000 km error free transmission has experimentally been achieved at 10 Gb/s in a recirculating loop. The evolution of Bit Error Rate for multiple pass into the regenerator and the polarization insensitivity demonstration to input data are presented.

## I. Introduction

Optical signal processing is now possible at high bit rate (up to 40 Gb/s) and could be an alternative way to all-electronic signal processing in high-capacity optical core networks. Signal regeneration is an attractive way in long haul optical data transmission. Indeed,, the signal undergoes several well-known impairments along propagation, like fiber attenuation and distortions due to chromatic dispersion and nonlinear effects, or the addition of noise due to EDFA. As a consequence, the signal must be regenerated before excessive degradation, in order to increase the transmission distance.

We can distinguish three schemes for optical regeneration.

Manuscript received 1st novembre, 2007.
This work was supported in part by the "Ministère de la recherche", the "Région Bretagne" and the European Founding for the Regional Development (FEDER).
V. Roncin, G. Girault, M. Gay, L. Bramerie and J.C.Simon are from CNRS FOTON-ENSSAT, University of Rennes, 6 rue de Kerampont, BP 80518, 22305 Lannion cedex – France (corresponding author phone: 33 (0)2 96 46 91 50; fax: 33 (0)2 96 37 01 99; e-mail: vincent.roncin@enssat.fr ).

The simplest one, called 1R (first R for Repeater), consists in a simple signal amplification. The 2R regeneration, for "Reshaping Repeater", allows the reduction of optical noise and the increase of optical data extinction ratio by use of an optical function with a fast nonlinear transmission. Finally, the 3R regeneration, for "Retiming Reshaping Repeater", is a complete regeneration for On-Off Keying (OOK) modulation signals including an optical clock extracted from input signal, which is then modulated by the data. The principle of the 3R regeneration is described in Fig. 1.

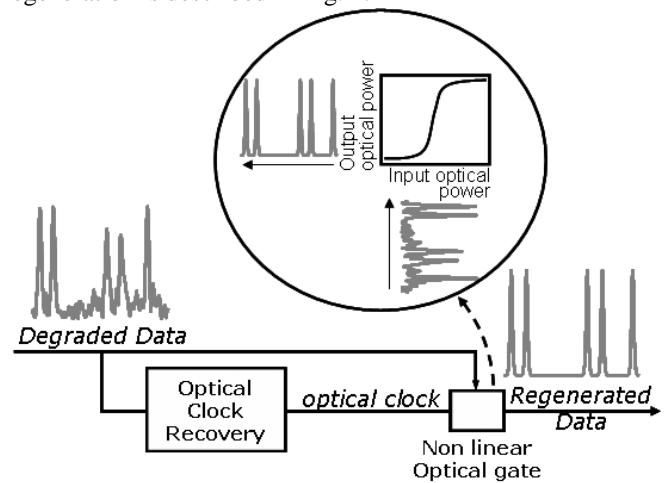

Fig. 1: 3R regeneration principle with nonlinear optical gate characterized by its nonlinear transfer function.

Degraded input data modulate a high-speed nonlinear optical gate which is able to reduce optical noise on data and transmitting an optical clock synchronised with the input data rhythm. Therefore, the jitter and the optical noise are both reduced. In order to achieve a complete regeneration, the useful nonlinear optical gate has to yield an S-shaped transfer function as shown on Fig. 1. Thus, thanks to its two nonlinearities, it reduces amplitude noise of the digital signal as well on symbols "1" as on symbols "0" [1], [2]. Furthermore, the higher its slope, the more the extinction ratio of regenerated data.

Several all-optical functions have already been studied as nonlinear optical gates for regeneration. Some of them using Self Phase Modulation (SPM) in a fibre associated with filter [3],[4], or a nonlinear interferometers [5], have been associated with synchronous modulation technique [6].

Thus, other highly reliable nonlinear components like Semiconductor Optical Amplifiers (SOA) have are very attractive. Actually, SOA have several advantages: compactness, possible polarization insensitivity, also low input power required for Cross Gain Modulation (XGM) or Cross Phase Modulation (XPM) operation. Thanks to their fast dynamic gain and their phase-amplitude coupling, SOA-based interferometers have been strongly studied achieving interesting results for optical regeneration [7], [8] and signal processing at high bit rate [9].

The Nonlinear Optical Loop Mirror (NOLM) is potentially a stable interferometer because the two delayed paths are in the same fiber. Therefore the fiber lengths, in which the co-propagative and contra-propagative fields propagate, are changing simultaneously, consequently reducing the phase difference noise (provided that the fiber is not too long). Its association with a semiconductor optical amplifier as the nonlinear element has often been proposed for all optical processing like demultiplexing in Optical Time Domain multiplexing (OTDM) context [10], [11],It has also been studied, to a less extent, for optical regeneration applications [12].

The SOA-NOLM is commonly used in transmission configuration which corresponds to "off-state" interferometer when no phase difference is applied between the two paths. In this paper we present a study of the SOA-NOLM in a reflective wavelength conversion configuration, which corresponds in this case to the "on-state" of the interferometer when no phase difference is applied between the two paths. The purpose of the proposed study is to evaluate the performance of this reflective configuration in the optical regeneration context.

In a first part, we describe the SOA-NOLM principle and we recall the main equations used in our numerical investigations. In a second part, we numerically study the transmission function of the interferometer in reflective configuration when the gain imbalance varies. In a third part we experimentally study and assess the low input polarization sensitivity of the SOA-NOLM. Finally we present the system assessment of a regeneration scheme combining this reflective SOA-NOLM wavelength converter, with a two-stage SOA wavelength converter

## II. SOA-NOLM BASIC EQUATIONS

The NOLM is an interferometer based on the Sagnac's loop set-up including a nonlinear device into its arm. As previously mentioned, the two waves which interfere propagate in the same length of fibres what confers to them lower sensibility to surrounding fluctuations (temperature and pressure) than other two waves interferometers like Mach-Zehnder and Michelson interferometers. Then, the phase response of this optical interferometer is obtained thanks to a phase-amplitude coupling into the nonlinear element. In the case of SOA-NOLM, the nonlinear element is an SOA in cross gain modulation configuration [13]. Many previous works also studied the capability of the NOLM for signal regeneration [14].

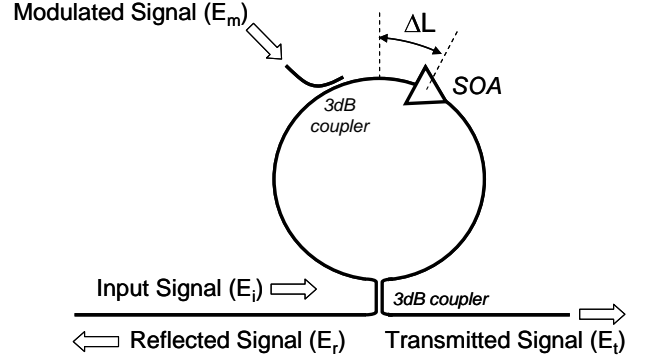

Fig. 2: SOA-NOLM operating principle. The SOA is the nonlinear element allowing to convert the input data signal into the transmitted or reflected output data signals

In Fig. 2, we present the SOA-NOLM operating principle. The input signal ($E_i$) is a continuous or a pulsed probe which is injected through a 3dB coupler and then divided into two co- and counter-propagating fields. The modulated pump signal ($E_m$), which is injected in the interferometer thanks to another 3dB coupler, corresponds to degraded data in the case of regeneration. It directly modulates the gain and as a consequence, the phase of the SOA thanks to the phase-amplitude coupling factor in semiconductor structures.

From the previous SOA-NOLM principles we then propose the basic matrix of the SOA-NOLM including well-know NOLM equations and amplitude modulation in the SOA [15]. The evolution of the reflected and transmitted fields $E_r$ and $E_t$ are defined as regard to the NOLM input theoretical 3dB coupling factor $K$, the SOA gain modulations induced by the co- and counter-optical powers $g_{co}(t)$ and $g_{counter}(t)$, and the corresponding non linear co- and counter-propagative optical phases $\varphi_{co}(t)$, $\varphi_{conter}(t)$ also in the SOA :

$$\begin{bmatrix} E_t \\ E_r \end{bmatrix} = \begin{bmatrix} \sqrt{1-K} & i\sqrt{K} \\ i\sqrt{K} & \sqrt{1-K} \end{bmatrix} \\ \times \begin{bmatrix} \sqrt{G_{co}} \cdot e^{-i\varphi_{co}} & 0 \\ 0 & \sqrt{G_{counter}} \cdot e^{-i\varphi_{counter}} \end{bmatrix} \quad (1) \\ \times \begin{bmatrix} \sqrt{1-K} & i\sqrt{K} \\ i\sqrt{K} & \sqrt{1-K} \end{bmatrix} \times \begin{bmatrix} E_i \\ 0 \end{bmatrix}$$

We deduce from it the power relations of the SOA-NOLM used in the numerical analysis. Moreover we simplify them by introducing the gain difference and the corresponding phase difference in the non linear interferometer:

$$\Delta G = \frac{G_{counter}}{G_{co}}$$
$$\Delta \varphi = \varphi_{co} - \varphi_{counter}$$

And we give the expression of the reflected and transmitted

optical power output of the SOA-NOLM:

$$P_T = P_i \cdot G_{co} \\ \left[(1-K)^2 + K^2 \Delta G - 2K(1-K)\sqrt{\Delta G} \cos(\Delta\varphi)\right] \\ P_R = K[1-K] \cdot P_i \cdot G_{co} \\ \left(1 + \Delta G + 2\sqrt{\Delta G} \cos(\Delta\varphi)\right)$$ (2)

The phase-amplitude coupling in the SOA is obtained thanks to the following relation with $\alpha_H$ the Henry factor in semiconductor structures.

$$\Delta\varphi = -\frac{\alpha_H}{2} \ln(\Delta G)$$

The gain modulation is caused by the input optical power $P_{in}$ such as $h$ the internal gain integrated through the SOA length can be expressed in a dynamic approach [16] with $P_{sat}$ the saturation power:

$$h(t) = \int_0^L g(z,t) \cdot dz = \int_0^L \frac{\partial P(z,t)}{P(z,t)} \cdot dz \\ \frac{\partial h(t)}{\partial t} = \frac{g_0 L - h(t)}{\tau_e} - \frac{P_{in}(t)}{\tau_e P_{Sat}}[\exp(h(t))-1]$$ (3)

The numerical values for each parameter used in the numerical study are summed-up in the following table.

| | |
|---|---|
| Wavelength (typical) | $\lambda = 1.55$ μm |
| Width (height of active section) | $w = d = 0.5$ μm |
| Effective area | $A_{eff} = 2.5 \, 10^{-13}$ m |
| Length | $L = 1 \, 10^{-3}$ m |
| Confinement factor | $\Gamma = 0.8$ |
| Gain coefficient | $a = 2.5 \, 10^{-16}$ cm$^2$ |
| Carrier density at transparency | $N_0 = 1.5 \, 10^{18}$ cm$^{-3}$ |
| Non radiative recombination coefficient | $A = 8 \, 10^8$ s$^{-1}$ |
| Radiative recombination coefficient | $B = 1 \, 10^{10}$ cm$^3$ s$^{-1}$ |
| Auger recombination coefficient | $C = 7 \, 10^{29}$ cm$^6$ s$^{-1}$ |
| Spontaneous recovering time of carrier density | $\tau_e = 0.5$ ns |

Tab. 1: Physical values used for the numerical simulation of the SOA

All the results of the SOA-NOLM numerical simulation in a static approach are obtained thanks to the stationary solutions of Eq. 3. We then use the following simplified expression in order to describe the static gain which only depends on the injected optical power and G0 the small signal gain of the SOA [17]:

$$G = \frac{G_0}{1 + \frac{P_{in}}{P_{Sat}}}$$

We can deduce the transfer function of the SOA-NOLM from its contrast C. We also introduce the extinction ratio parameter *ER* of the SOA-NOLM output signal which is often used in digital transmission studies to evaluate the quality of a digital signal. In our study, it allows to describe both static and dynamic behaviour

$$C = 1 - ER \\ ER = \frac{I_{max} - I_{noise}}{I_{min} - I_{noise}}$$

$I_{noise}$ is the optical noise intensity generated by the SOA in the interferometer, which limits the contrast of the modulation. $I_{max}$ and $I_{min}$ are deduced from Eq. (2).

In fig. 3, the contrast of the SOA-NOLM output power is plotted, respectively for configuration in transmission (*It*) and reflection (*Ir*) The maximum of contrast is obtained for each configuration with a phase difference of $\Delta\varphi_{SOA} = \pi$. For the reflective power *Ir*, the theoretical maximum is about 13 dB and 30 dB for the configuration in transmission (*It*). For the theoretical result, it is important to note that the optical spontaneous emission continuous power $P_{opt\,noise}$ is -30 dBm.

So the SOA-NOLM in reflection configuration presents a lower output contrast but a higher robustness to signal degradation and external instabilities, which often are the main problem of optical functions based on interferometer. Indeed we observe that the slope of the reflected contrast presents firstly, two saturation zones allowing to reduce, the noise fluctuations on "1" and "0" levels and secondly, a real "S" shape which is required for multi-pass optical regeneration.

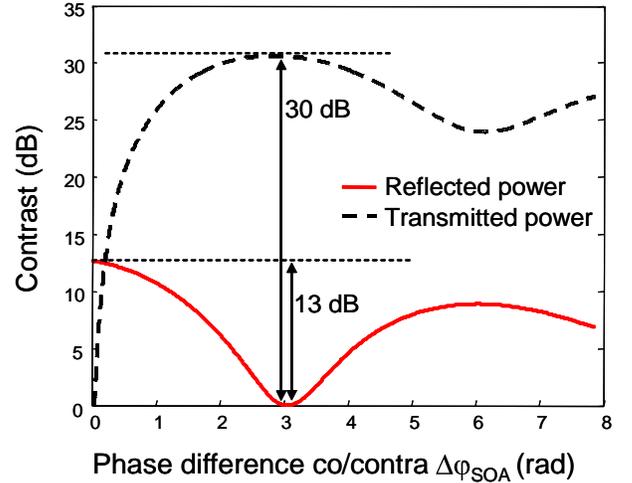

Fig. 3: Theoretical SOA-NOLM contrast evolution for the two considered configurations. Effect of the differential phase shift $\Delta\varphi_{SOA}$ corresponding to the co- and counter-propagative probe phase difference.

III. NUMERICAL STUDY OF THE SOA-NOLM IN REFLECTIVE CONFIGURATION

In this study we introduce the static transfer function which represents the evolution of the probe power versus the input pump power. This function $f_{sat}$ is determined in the case of continuous probe ($P^{probe}$) and pump ($P^{pump}$). We use the following expression:

$$\frac{P_{out}^{probe}}{P_{in}^{probe}} = f_{stat}\left(P_{in}^{pump}\right)$$

For the temporal analysis of the optical function, we introduce the dynamic transfer function, $f_{dyn}$, which is obtained with a modulated pump signal defined temporally. The function gives us the relation between the input power $P^{in}$ and the output probe contrast, $C^{out}$ defined from (16). In the following expression, $f_{dyn}$ is obtained from the equations of the dynamic gain.

$$C_{out}^{probe} = f_{dyn}\left(P_{in}^{pump}\right)$$

Hereafter, we study and compared the transfer functions of the SOA-NOLM in the static and dynamic approach. Then, we numerically check that both approaches lead to similar results and that the SOA-NOLM dynamic transfer function using $\Delta G_{SOA} = 1$, corresponds to the static transfer function of a single SOA.

Moreover, in this study we define, two different origins which can generate a gain difference in the SOA-NOLM:

The first one is linked to the static approach and corresponds to the small-signal gain difference of the co-propagative probe and the counter-propagative one. We talk about the static gain difference $\Delta G_{SOA\text{-static}}$, caused principally by imbalanced losses in the set-up, between the co and the counter-propagating optical ways.

The second one is the instantaneous gain difference caused by the position of the SOA in the NOLM being able to cause a temporal delay, already mentioned, between the co and counter-propagating optical ways.

In the following paragraph, we use the previously described equations to study the SOA-NOLM transfer function.

*A. Study of the static transfer function*

In that case, the pump and the probe are continuous waves. Consequently, there are no phase changes generated by the position of the SOA in the NOLM. Thus the only phase difference is caused by the static gain difference between the co- and counter-propagating gains induced by the loss imbalance of the set-up. This effect can be understood if the co and counter injected probe powers are themselves imbalanced. Therefore, a modification of the optical saturation power is introduced for the co and contra optical gains in the SOA. This generates then a static gain difference.

We theoretically study the transfer function by using the following parameters: a continuous probe $P_{probe} = -5$ dBm, a linewidth enhancement factor $\alpha = 7$ [18] and a bias current of 300 mA for the SOA. Considering a classical parabolic gain evolution with the wavelength, we do not consider any gain dependency, in our study, induced by the wavelength shift between the pump and the probe.

The SOA-NOLM equations are given for a coupling coefficient of the input 3 dB coupler of 0.45. We change the value of $\Delta G_{SOA\text{-static}}$, the static gain difference between co and counter propagative fields, in the range of 1.6 to 2.6 dB in order to describe the inflection of the transfer function. This parameter includes the loss imbalance of the set-up. The range of $\Delta G_{SOA\text{-static}}$ is chosen in order to lead to a sufficient SOA-NOLM static transfer function transformation and a modification of its slope while remaining in a realistic configuration. It is important to note that this static phase difference does not induce any temporal distortion on the output pulses. This point is important in the case of multi-pass into the SOA-NOLM.

Theoretical results are presented on the Fig 4. For this unbalanced configuration, we clearly observe a slope increase for the static transfer function of the SOA-NOLM with the increase of $\Delta G_{SOA\text{-static}}$ and as compared to the SOA alone. Therefore, the best transfer function for optical regeneration is observed for a value of 2.4 dB for $\Delta G_{SOA\text{-static}}$. Moreover, we observe a contrast of the SOA-NOLM higher than 9 dB for an optical input power of about 5 dBm. In the following paragraph we study the impact of the SOA position in the NOLM in a dynamic configuration. Non linearity and contrast of the SOA-NOLM transfer function will then be studied.

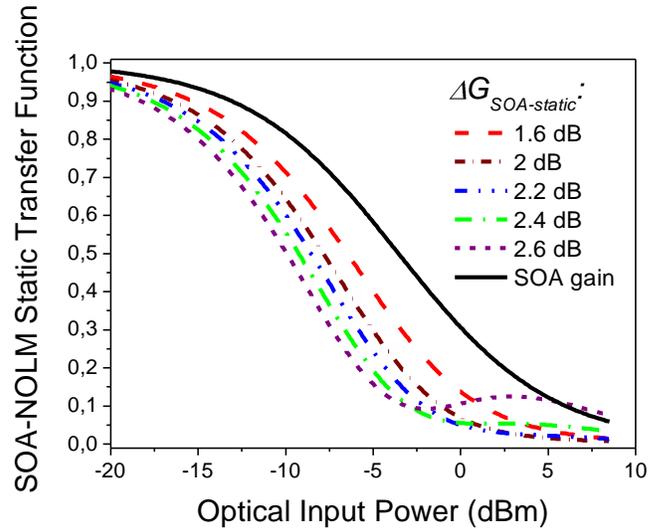

Fig 4: Theoretical comparison between the static transfer function of the lonely SOA and the SOA-NOLM. Improvement of the nonlinear shape for the SOA-NOLM, induced by the static co/counter gain difference $\Delta G_{SOA\text{-static}}$

*B. Study of the dynamic transfer function*

In this part we theoretically study the influence of the position of the SOA in the SOA-NOLM architecture. This configuration introduces a temporal imbalance which is well-known in optical switching application because it allows to open a temporal window corresponding to the delay on the arrival of the probe pulses into the SOA [19]. It is then possible to create a temporal gain difference in the SOA, previously call $\Delta G_{SOA\text{-dynamic}}$. In the previously proposed Tab. 1, we have presented all the physical values computed for theoretical simulation of SOA and SOA-NOLM. The obtained results have been successfully compared with another quantitative model [20] with 20 ps pulses-width at 10 GHz.

For the numerical simulation, the optical input is an RZ 10 Gb/s signal with 20 ps pulse-width. The extinction ratio of the signal is 25 dB. The SOA bias current is 300 mA and the input power of the continuous probe is −10 dBm. The signal is presented in Fig. 5.

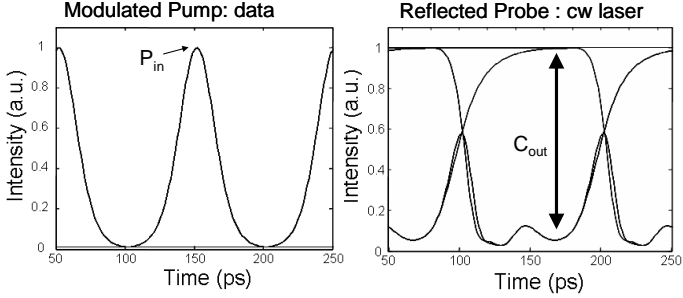

Fig. 5: Signal characteristics for the simulation with injected data (ER=25 dB) on the left and reflected probe (modulated by the pump) on the right. The delay used for the simulation, introduced by the SOA position in the SOA-NOLM is 15 ps

On the one hand, we deduce the Dynamic Transfer Function (DTF) from the dynamic contrast of the reflected output of the SOA-NOLM. The characteristics of the signals used for the simulation are presented in Fig. 5. The critical parameters are the optical input power of the input data Pin and the output probe contrast $C_{out}$. We then deduce the DTF by varying the optical input power, and this for different delays caused by the position of the SOA in the interferometer. The results are reported in Fig. 6 (a).

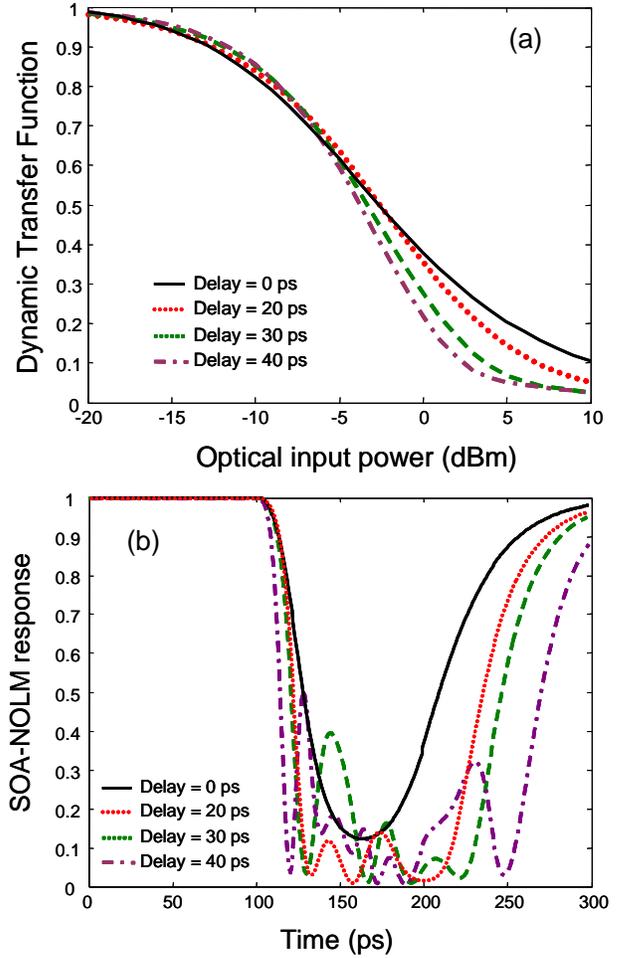

Fig 6 : (a) Theoretical dynamic transfer function of the SOA-NOLM for different delay between co and counter probe. (b) Theoretical temporal response of the SOA-NOLM for an input [0 1 0] sequence for the same delays in the SOA-NOLM Input signal is a 20 ps RZ at 10 Gb/s.

On the other hand, we present the temporal distortions caused by the delay in the SOA-NOLM. The results of the simulation are presented on Fig.6 (b).

In fig 6-(a), we show the modification of the transfer function of the SOA-NOLM according to the delay linked to the position of the SOA, which modifies the $\Delta G_{SOA\text{-dynamic}}$ and thus the $\Delta \varphi_{SOA}$. We observe that, for a delay of 0 ps, the transfer function of the SOA-NOLM and the SOA are the same. For other values, an instantaneous phase difference is generated leading to the modification of the slope of the transfer function. Thus the more important the delay is, the more the dynamic transfer function slope increases.

The steepest transfer function is obtained for a delay value of 40 ps. Therefore, for an input extinction ratio of 25 dB, the output extinction ratio is more than 15 dB and the "S" shape is clearly demonstrated, despite the extinction ratio degradation. This result shows that the delay allows to shape the DTF in order to increase the regenerative capacity of the SOA-NOLM in a reflective scheme. However, we can not disconnect the transfer function to the signal quality and particularly to the temporal distortions induced on the optical signal. It is the reason why we limit the delay to 40 ps at a bit rate of 10 Gb/s.

For higher delays, the distortions are not compatible with 3R applications anymore. This point is clearly shown in Figure 6-(b). These traces present the temporal response of the SOA-NOLM with an input data signal corresponding to a RZ 50% at 10 Gb/s, with a binary sequence of [0 1 0]. We observe that the SOA-NOLM response is similar to the SOA gain compression if the delay is 0 ps. For other values of delays, the compression is increased but temporal distortions appear. So, the higher the delay is, the more important the distortions are, which can lead to dramatic patterning effects in a real transmission system. In the experimental part we have tried to increase the value for the delay, and also to minimize the distortion on the SOA-NOLM output modulation.

## IV. EXPERIMENTAL RESULTS

We have theoretically shown that a SOA-NOLM in reflective configuration is a suitable optical function for all-optical regeneration at 10 Gb/s. We are now going to experimentally study its ability to regenerate signals. In a first paragraph, we show that the function is polarization insensitive and in another one, we evaluate its qualities as a regenerator in a system environment.

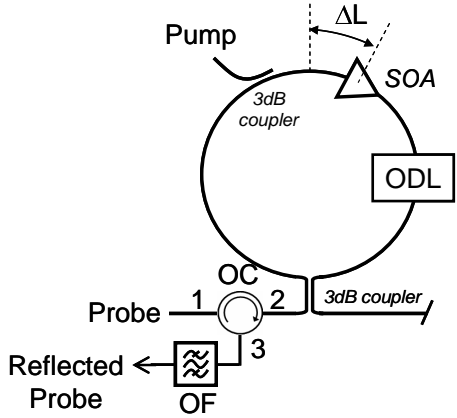

Fig. 7: Scheme of the SOA-NOLM experimental set up in using the reflective port.

The experimental set up of the polarization maintaining SOA-NOLM is made up with an optical fibre is presented on Fig. . An optical circulator (OC) allows to collect the reflected signal. The port "1" of the circulator is connected to the probe input, the port "2" to the SOA-NOLM and the port "3" allows to extract the output reflected signal from the SOA-NOLM. This circulator is followed by an optical filter (OF) for rejecting the pump signal. The optical distance ΔL between the SOA and the middle of the fiber loop is adjusted with an optical delay line (ODL). The SOA has been designed by Alcatel CIT for 10 Gb/s applications. Its small-signal gain is up to 30 dB and centred around 1550 nm. The laser is packaged in a module to ensure the efficiency of the coupling into the fibres. This coupling is free-space mounted for maintaining the optical polarization in the set-up.

### A. Polarization sensitivity analysis

The polarization insensitivity of a device is a fundamental property for an optical function in a transmission system environment [21]. Indeed, optical polarization of telecommunication signals at the entrance of an optical function is a random parameter which changes continuously. The characteristic parameter generally used for the polarization sensitivity estimation of a device is the Polarization Dependant Loss (PDL). For active functions like optical amplifiers, the Polarization Dependant Gain (PDG), which corresponds to the amplification gain sensitivity induced by the input signal polarization, is used. The PDG is defined as the difference, in decibel, between the two transfer functions obtained with a linear polarized signal, according to the two main axes of the device. These axes are commonly called TE and TM corresponding to the TE and TM main propagative modes in the SOA waveguide [22]. Then it is possible to plot the two gain functions of the amplifier, respectively called $G_{TE\_dB}$ and $G_{TM\_dB}$. Thus, the PDG is calculated thanks to the following expression:

$$PDG_{dB} = \left| G_{TE\_dB} - G_{TM\_dB} \right| \quad (1)$$

In our configuration, the PDG of the SOA and the SOA-NOLM (in reflected configuration) are thus obtained considering the gain $G_{co\_dB}$ measured for the co-propagative field and $I_{r\_dB}$, the reflected intensity at the output of the SOA-NOLM.

The experimental set up used to measure the polarization sensitivity of the devices is described on Fig. 8. In order to alternatively impose the TE or TM polarizations of the pump we have used a polarization controller to adjust the state of polarization in front of a variable polarizer (VP). We ensure that the polarization is stable if the power level after the VP is a maximum. The linear polarization at the SOA input is checked, before each measurement, thanks to a polarimeter (Profile PAT 9000). Each measurement has been made several times and the measurement error has been assessed to 0.3 dB. This value corresponds to the PDL of the set up without DUT (Device Under Test). Also in Fig. 8, we present the two DUT with the port (1) for the pump, the port (2) for the probe and the port (3) for the output.

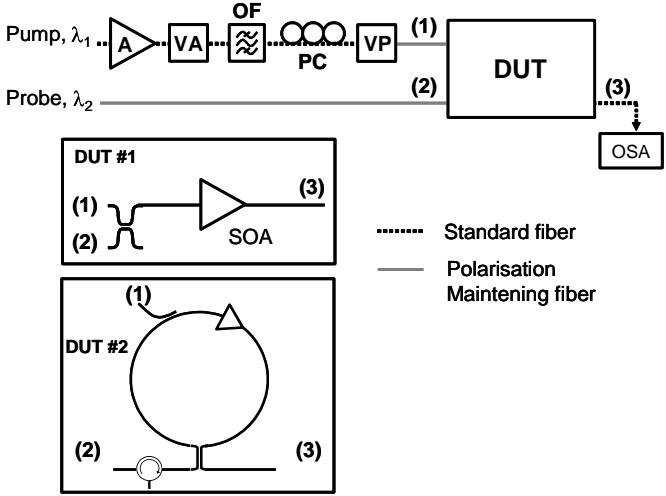

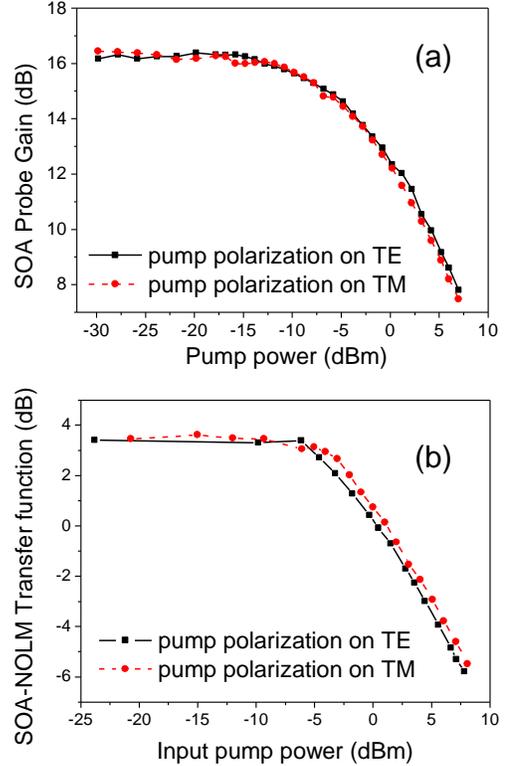

Fig. 8: Experimental set-ups for polarization sensitivity analysis of a DUT in pump-probe configuration: DUT#1 is a single SOA and DUT#2 is the SOA-NOLM in reflection scheme. "A" is an optical Amplifier, "VA" a Variable Attenuator, "OF" a bandpass Optical Filter for OSNR improvement, "PC" a Polarization Controller and "VP" a Variable linear Polarizer.

The SOA in the NOLM is used in cross-gain modulation with high pump power levels. Therefore, the SOA PDG must be compared to SOA-NOLM PDG in this configuration. The results are presented in Fig. 9-(a) for the SOA alone and in Fig. 9-(b) for the SOA-NOLM.

For a better comprehension of the set-up, we notice that the SOA-NOLM has been built in Polarization Maintaining Fibres and that the probe is linearly polarized thanks to a polarized optical source. The probe polarization is preserved in all the interferometer and particularly at each side of the SOA. For the pump polarization, we choose the VP position to reach the TM or the TE linear polarization corresponding to a maximum and a minimum of the small-signal gain. The other polarization cases correspond to a combination of TE and TM polarizations.

The curves shown on Fig. 9 allow experimentally to deduce the PDG of each device by subtracting the square and dot curves. We thus obtained an experimental PDG equalled to 0.5 dB for the SOA as well as the SOA-NOLM, whatever the pump power. So the SOA and the SOA-NOLM are totally polarization insensitive in this case of operating. Further investigations should be necessary to clearly explain the polarization insensitivity of the SOA. This result has been obtained with a probe signal at 1557 nm and a pump signal at 1552 nm. However, we can note that the polarization insensitivity of the SOA-NOLM is directly linked to the SOA polarization insensitivity in the studied pump-probe configuration.

So we experimentally demonstrate that the SOA-NOLM is potentially polarization independent in wavelength-conversion, which is a necessary condition for 3R regeneration application in a system application.

Fig. 9: Experimental transfer functions of the single SOA (a) and the SOA-NOLM (b) in pump / probe configuration. These results allow the calculation of the polarization dependant gain (PDG) of each device.

### B. Regeneration Performance in system

As previously shown in the numerical analysis, the SOA-NOLM has a nonlinear transfer function, suitable for regenerative applications. We have also experimentally shown the polarization insensitivity of the SOA-NOLM, when made of PM fibre, in cross-gain modulation and reflective configuration.

In this part, we propose to demonstrate the optical regeneration capability of the SOA-NOLM in a system experiment which necessarily preserves the wavelength and the data polarity [23].

Therefore, we use a second wavelength-converter and also data polarity inverter which has been proposed hereafter to increase the output extinction-ratio without particular regenerative properties: the Dual-Stage of SOA (DS-SOA) [24]. The main advantage of the DS-SOA is to increase the extinction ratio twice, in decibel, as compared to a single SOA.

The complete set-up used for the 3R optical regenerator experiment, is presented in Fig 10. The input signal is a 10 Gb/s data, with RZ 50% format, at $\lambda_1$=1552 nm. The second optical signal is a linearly polarized continuous laser, at $\lambda_2$=1557 nm, which is injected into the SOA-NOLM. The polarization of the continuous laser is collinear to the TM axis of the SOA. The reflected output signal is converted at the wavelength $\lambda_2$ and data polarity inverted, as shown in the eye diagram inset of Fig.10.

The SOA-NOLM output data signal is amplified and next injected into each SOA of the DS-SOA to modulate simultaneously their gains. The optical clock used for the 3R configuration is obtained with a commercial optoelectronic clock recovery. The electrical signal modulates a continuous laser at $\lambda_1$=1552 nm, through an optical modulator. It is injected as a probe signal in both SOA, to be modulated twice by cross-gain modulation (XGM).

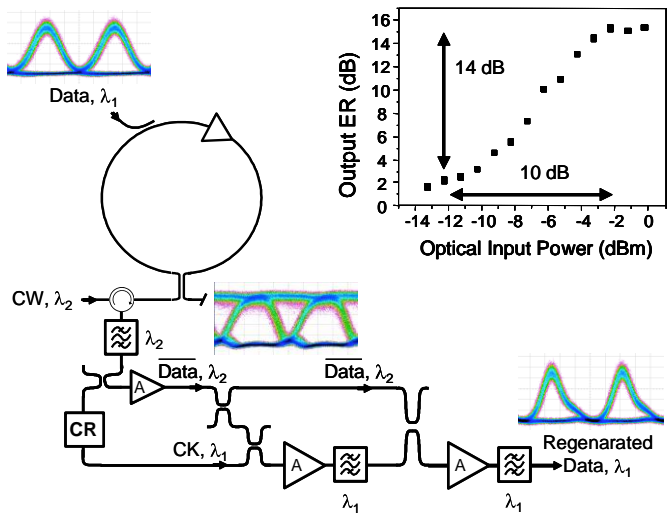

Fig. 10: 3R regenerator scheme with "A" for EDFA and "CR" for optoelectronic Clock Recovery, "CK" for optical clock. The curve in inset is the dynamic transfer function of the complete regenerator obtained from data extinction ratio measurement with a sampling oscilloscope

We have experimentally measured the transfer function of the complete 3R regenerator. The curve is plotted in the inset of Fig 10. The measurement has been succeeded by reducing the power of the input data and then measuring the output regenerated data with a sampling oscilloscope. The S-shape is clearly shown and an extinction ratio of 10 dB for the input signal (input power) corresponds to an output extinction ratio of more than 14 dB. This extinction ratio improvement combining to the "S" shape ensure a full 3R regeneration.

In the transmission system experiment, we used an optical power for data signal with the level of symbols "0" lower than -12 dBm, and level of symbols "1" higher than -3 dBm. This configuration ensures an efficient operating of the regenerative function of the SOA-NOLM. Therefore, the optical noise on symbols "0" and symbols "1" is reduced. Concerning the retiming process, which corresponds to an efficient timing jitter reduction, the recovered optical clock has been chosen to be narrower than the timing window of the regenerated data.

The best way to validate a regenerator is to test it in a transmission link [25]. So we test the efficiency of the regeneration function composed by the SOA-NOLM in a recirculation loop. This experimental set-up allows to do multiple pass into the regenerator and then, to accumulate the effect of regeneration, and also to observe and quantify its efficiency.

The experimental set up of the recirculation loop is represented on Fig. 11-(a). The loop is composed of two 50 km Non-Zero Dispersion Shifted Fiber (NZDSF) spans, with chromatic dispersion of 4ps/nm/km. The fiber link dispersion is compensated (DCF).

Losses are compensated by Erbium Doped Fibre Amplifiers (EDFA) and counter-propagating Raman pumping ensuring a low noise accumulation line. The transmitter consists of a $2^{15}-1$ pseudo-random bit sequence combined with a logical gate which produces an RZ electrical signal. This signal modulates the optical 1552 nm source thanks to a $LiNbO_3$ modulator which produces a 50 ps full width at half maximum signal. The signal is injected into the recirculating loop thanks to an Acousto-Optic Modulators (AOM).

A polarization scrambler (polarization modulation frequency ~ 1 MHz) is placed in front of the regenerator in order to take polarization effects into account and to check the stability of the device versus polarization fluctuations and highlight the polarization insensitivity of the 3R regenerator.

With the previously described set up, without regenerator, a maximal transmission of 2000 km has been achieved. When including in the loop the 3R device composed by the SOA-NOLM followed by the DS-SOA, an error free transmission of 100 000 km, corresponding to 1000 laps of 100 km, has been achieved. This result has been obtained for an OSNR (Optical Signal to Noise Ratio) in the loop, at the input of the 3R regenerator, of 33 dB / 0.1 nm.

Then in order to do BER (Bit Error Rate) measurements, noise is artificially included in the loop, using an Amplified Spontaneous Emission source (ASE). This allows degrading the OSNR in front of the regenerator which is necessary to measure a BER in regenerated signal experiments [26]. The evolution of the BER versus the number of laps in the loop, which corresponds to the transmission length, can thus be plotted. The results are reported on Fig.11-(b). The empty and full symbols respectively correspond to tranmissions without and with the 3R regenerator. The curve with square symbols has been obtained with an OSNR of 24 dB in front of the regenerator and the curve with triangles with an OSNR of 21 dB.

These curves show that the tranmission length is significantly improved when including the 3R regenerator in the loop. Moreover a transmission with a BER of $10^{-8}$ is obtained with an OSNR of 24 dB measured in front of the regenerator. With an OSNR of 21 dB a transmission with a BER of $10^{-4}$ is achieved which is a good performance considering the possibility of using a FEC (Forward Error Correcting Code). We can note that the shape of the regenerated curve corresponds to the classical shape preview for 3R regenerated transmission with optimal regenerator.

We can finally note that all these measurements have been achieved thanks to the high stability of the proposed regenerator, widely studied and commented in the article.

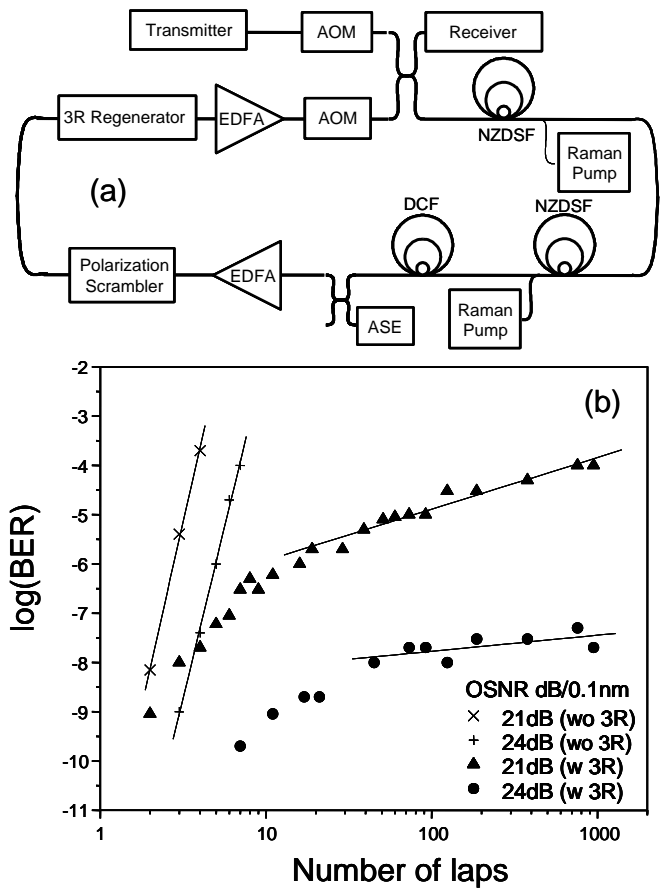

Fig. 11: Experimental set-up of the 100 km length recirculating loop (a), BER measurements in recirculating loop experiments (b) for non-regenerated transmission (without 3R) and regenerated transmission (with 3R).

## V. CONCLUSION

In this work, we studied, in an original configuration, a classical optical function often used for optical signal processing: the SOA-NOLM in a reflective scheme. The interferometer used in this configuration is a wavelength converter which changes the data polarity. In this complete article, we demonstrated experimentally that the SOA-NOLM is very stable, thanks to its intrinsic properties linked to its singular construction: one single arm and one unique SOA, which is in this configuration, polarization insensitive. We then used its capability of nonlinear transmission and its properties for optical regeneration operation. We theoretically studied its transfer function and highlighted two different ways to increase its slope, as it is required in the optical regeneration theory. We shew the results obtained by changing dynamically and statically the phase difference between the co and counter-propagative fields in the interferometer, thanks to the phase-amplitude coupling in the SOA.

Finally, we tested this function experimentally by showing in a first paragraph its optical polarization insensitivity and consequently, in a second part its capability to work in an data transmission environment. So we present the result of a 3R regenerated transmission of 100 000 km, without detecting any error.

## VI. ACKNOWLEDGMENT

The authors would like to thank all the team "Fonctions Optiques pour les Télécommunication" for the fruitful discussions. This work has been supported by the European Found for Regional Development FEDER; the "Ministère de la recherche" and the "Region Bretagne".

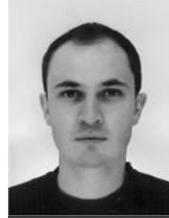

**Vincent Roncin** received the Master degree in "lasers, communications and metrology" from the University Paris XIII, France, in 2000; and the Ph.D degree in Physics from the University of Rennes, France, in 2004. He is now researcher in optics at CNRS-FOTON and the manager of two projects on optical clock recovery using passive and active techniques. His background is within optical regeneration, optical signal processing based on SOA and optical clock recovery until 160Gbit/s. Its recent interest is in phase and intensity optical noise, in self-pulsating lasers based on quantum-dot and bulk structures.

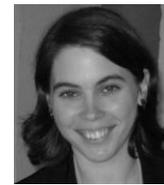

**Gwenaëlle Girault** received the Engineer degree in optronics in 2003, the Masterdegree (D.E.A.) in sciences and techniques of communication the same year, and the PhD degree in optical communications and physics in 2007 from CNRS unity FOTON-ENSSAT, University of Rennes I, France. The main topic of her research is the study of semiconductor optical amplifier based optical functions for all-optical regeneration and optical system transmission. She is now research engineer on PERSYST platform, independent public research and test facility offering a testbed for 40Gb/s and 10Gb/s optical telecommunications systems open to private companies and university teams.

**Mathilde Gay** received her PhD degree in physics in 2006 from INSA, France. She works now as a research engineer at CNRS FOTON-ENSSAT on the PERSYST platform. Her background is within the theoretical and experimental investigation of the impact of optical regeneration on high bit rate optical transmission systems. She has experience as characterisation platform manager within ePIXnet network of excellence.

**Laurent Bramerie** received the opto-electronic engineering degree from ENSSAT, University of Rennes I, France, in 1999 and the PhD degree in 2004. He worked, France as technical expert on ultra-long haul 40 Gb/s DWDM systems for 2 years in Corvis Algety in Lannion, France. In 2003 he has joined CNRS FOTON-ENSSAT where is now research engineer on PERSYST platform, independent public research and test facility offering a testbed for 40Gb/s and 10Gb/s optical telecommunications systems open to private companies and university teams.

**Jean-Claude Simon** received the Doctorat d'Etat degree from Université de Nice in 1983. From 1975 to 1998 he was with CNET, the research centre of the French PTT (now France Telecom R&D) as a researcher in the field of semiconductor optical amplifiers and non-linear optical signal processing, principally 2R and 3R all-optical regeneration. In 1999 he moved to ENSSAT, an engineering school, as a full professor in optoelectronics. He is director of Laboratoire d'Optronique, a research lab. associated with CNRS and Université de Rennes I. He has authored or co-authored approximately 160 journal and conference papers including some 25 invited presentations, 10 patents, and 3 book chapters. He has contributed to European research programs (RACE 1027, ESPRIT 3 MOSAIC and ACTS KEOPS). He is recipient of the Fabry-de Grammont award from the French Optical Society.